\documentclass[twocolumn,aps,prl,showpacs]{revtex4}
\usepackage{graphics,amsmath,graphicx}

\begin{document}

\title{Transient localization 
in crystalline organic semiconductors.}

\author{S. Ciuchi$^1$,  S. Fratini$^2$ and D. Mayou$^2$}

\affiliation{$^1$Istituto dei Sistemi Complessi CNR, CNISM and Dipartimento di Fisica,
Universit\`a dell'Aquila, via Vetoio, I-67100 Coppito-L'Aquila, Italy\\
$^2$ Institut N\'eel-CNRS and Universit\'e Joseph Fourier,Bo\^ite
Postale 166, F-38042 Grenoble Cedex 9, France}

\date{\today}

\begin{abstract}
A relation derived from the Kubo formula shows that 
optical conductivity measurements 
below the gap frequency in doped semiconductors 
can be used to probe directly the 
time-dependent quantum dynamics of charge carriers.
This allows to extract fundamental quantities such as the
elastic and  inelastic scattering rates, as well as the
localization length in disordered systems.
When applied to crystalline organic
semiconductors, an incipient electron localization
caused by large dynamical lattice disorder is unveiled, 
implying a breakdown of semiclassical transport.
\end{abstract}

\maketitle

\paragraph{Introduction.}

``Bad'' conductors are systems presenting
a breakdown of
the semiclassical Bloch-Boltzmann
description of electronic transport. Known examples are
found in various
classes of materials such as disordered systems 
\cite{LeeRMP}, transition metal 
compounds \cite{GunnarssonRMP},
alkali-doped and charge-transfer organic metals \cite{theta-I3} and
quasi-crystals \cite{Mayou00,Trambly06}.
In all these cases, the electron motion is so much slowed down
(by disorder, electronic correlations, polaronic effects, or by
structural constraints)
that the semiclassical assumption of well-defined wave-packets undergoing
rare scattering events is not valid.

An analogous situation is encountered in crystalline organic
semiconductors. There, a fundamental unsettled question
is whether the mechanism of charge transport can be ultimately
understood from the point of view of band electrons alone, as
suggested by the ``band-like'' temperature dependence of the measured mobility.
In these materials, the inherently large thermal molecular motions
act as strong electron scatterers,
leading
  to apparent electron mean-free-paths comparable or even smaller than the
intermolecular distances.
While a generalization of semiclassical
transport theory has been recently proposed in Ref. \onlinecite{Fratini09}
to deal with this situation,
it might well be that the very nature of semiclassical transport is
insufficient to appropriately describe the charge transport mechanism
in these materials. Indeed, recent numerical studies have suggested
a radically different point of view, based on
a form of electron localization due to the dynamical
disorder caused by the thermal  molecular motion
 \cite{Troisi,Picon}.
Accordingly, a theory of electron transport in organic semiconductors would
require a proper account of quantum corrections to the
electron dynamics, 
not included in semiclassical treatments.

In this paper we tackle this problem by expressing 
the Kubo formula as a relation between 
the optical conductivity and the time-resolved quantum dynamics of
electrons.  We first consider a microscopic model with dynamical
lattice disorder  that illustrates the characteristic behavior of the quantum
diffusion. A relaxation time approximation is then introduced
that treats the effect of inelastic scattering by low-frequency lattice
vibrations in an intuitive way.
Finally these concepts are  used  in the interpretation of
experimental data in crystalline rubrene, providing evidence for
localization effects.


\paragraph{Formalism.}
The quantum diffusion of
electrons in a given spatial direction
can be measured via their quantum-mechanical spread
\begin{equation}
   \label{eq:X2}
   \Delta X^2(t)=\langle [ \hat X(t) - \hat X(0) ]^2 \rangle,
\end{equation}
where $\hat X(t)=\sum_{i=1}^N \hat x_i(t)$
is  the total position operator of $N$ electrons in the Heisenberg
representation and $\langle \cdots \rangle= \mathrm{Tr}[ e^{-\beta H}
(\ldots)]/Z$ denotes the thermodynamic average.
$\Delta X^2(t)$ is directly related to the {\it symmetrized}
self-correlation function  of the velocity operator $\hat V_X(t)=\frac{d\hat
   X(t)}{dt}$,  $C(t)= \langle \hat V_X(t) \hat V_X(0) + \hat V_X(0)
\hat V_X(t)\rangle$, via \cite{Mayou00,Trambly06}
\begin{equation}
   \label{eq:vv}
   \frac{d \Delta X^2(t)}{dt}=
  \int_0^t C(t^\prime) dt^\prime.
\end{equation}
On the other hand, the Kubo formula expresses the dissipative
part of the  optical conductivity as
\begin{equation}
   \label{eq:Kubo}
   \sigma(\omega) = \frac{e^2}{\nu\hbar\omega} Re \int_0^\infty  e^{i\omega t}
   \langle [\hat V_X(t),\hat V_X(0)] \rangle dt,
\end{equation}
where $\nu$ is the volume of the system.
This can be exactly
related to the  symmetrized $C(t)$ of Eq. (\ref{eq:vv}), and
therefore, to the quantum diffusion  $\Delta X^2(t)$.
Replacing the commutator  in
Eq. (\ref{eq:Kubo}) with the anticommutator of Eq. (\ref{eq:vv})
can be absorbed into a 
detailed-balance
prefactor, yielding:
\begin{equation}
   \label{eq:relation}
\sigma(\omega)=-\frac{e^2\omega^2}{\nu}\frac{\tanh 
(\beta\hbar\omega/2)}{\hbar\omega}
   Re \int_0^\infty e^{i\omega t} \Delta X^2(t) dt
\end{equation}
with $\beta=1/k_BT$ (see 
Refs.\onlinecite{Lindner10,Kubo} for related aspects).
The above Eq. (\ref{eq:relation}) is a restatement of the Kubo formula,
identifying the time-dependent
quantum diffusion as the physical quantity that is dual
to the optical absorption in the frequency domain. 
For independent non-degenerate
electrons, the formalism presented above
acquires an  intuitive meaning in terms of the quantum spread of the electronic
wavefunctions, as in this case one has
$\Delta X^2(t)=N \Delta x^2(t)$, with $\Delta
x^2(t)$ referring to each individual particle.
Eq. (\ref{eq:relation}) can be inverted  to give
\begin{equation}
   \label{eq:relinv}
  \Delta x^2(t)=-\frac{2 \hbar}{\pi e^2}
  Re \int_0^\infty e^{-i\omega t}\frac{ 
\sigma(\omega)/n}{\omega\tanh 
(\beta\hbar\omega/2)}
    d\omega.
\end{equation}
with $n=N/\nu$ the electron density.

\begin{figure}
   \centering
   \includegraphics[width=8.5cm]{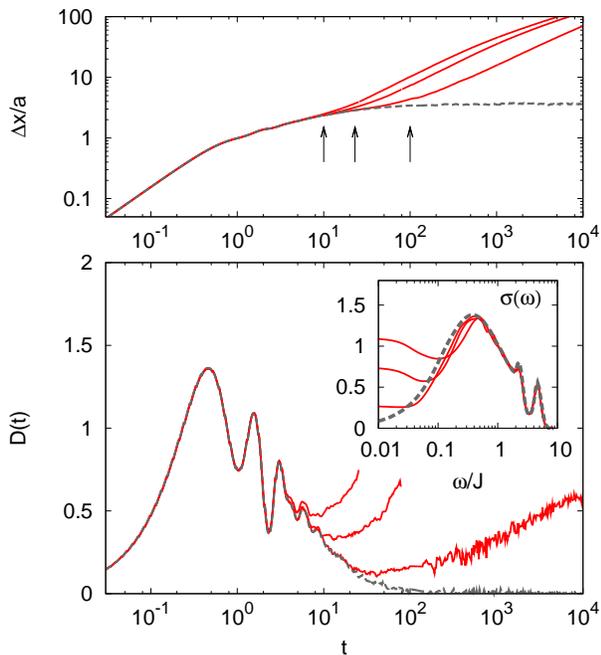}
   \caption{\label{fig:theory}
     a) Quantum spread obtained from the microscopic
     model Eq. (\ref{eq:SSH}) for static  (grey,
     dashed)  as well as dynamical  disorder (red, full lines: from bottom
     to top, $\hbar \omega_0/J=0.01,0.0435,0.1$). Times
     are in units of $\hbar/J$. b)
The corresponding  instantaneous diffusivity $D(t)=(d\Delta x^2/dt)/2$.
The inset shows the optical conductivity
     obtained via Eq. (\ref{eq:relation}).
  }
\end{figure}

\paragraph{Microscopic description of lattice disorder.}

To set the concepts clear, we apply the quantum diffusion
formalism to the following Hamiltonian \cite{Troisi,Fratini09},
\begin{equation}
   H= - J\sum_i  [1-\alpha(u_i-u_{i+1})] \; (c^+_i  c_{i+1} + c^+_{i+1}
c_{i}) + H_{ph}.
   \label{eq:SSH}
\end{equation}
In this model, electrons in a one-dimensional
tight-binding band have their inter-molecular transfer integrals
modulated by
molecular vibrations of frequency $\omega_0$, represented by
  $H_{ph}= \sum_i  \frac{M \omega_0^2u_i^2}{2}
  + \frac{p_i^2}{2M}$. Eq. (\ref{eq:SSH})  captures
the effects of strong dynamical lattice disorder that seem to be crucial
in  crystalline organic semiconductors.
The phonon frequency $\omega_0$ is
  small due to the large molecular weight, so that
  the  lattice fluctuations can be treated classically 
  ($k_BT > \hbar \omega_0$). 
   Their  coupling to electrons is governed by the dimensionless parameter
$\lambda=\alpha^2 J/(2 M\omega^2_0)$.

To calculate the electron 
diffusion $\Delta x^2(t)$ in the presence of lattice dynamics we
employ mixed quantum-classical simulations based on the Ehrenfest
coupled equations \cite{Troisi} on a $1024$-site chain.
We solve the Schr\"odinger equation and average over up to
$12800$ initial conditions, with the initial displacements, ${u_i}$,
obeying a thermal distribution $P(u_i)\propto
\exp(-M\omega_0^2u_i^2/2k_B T)$.
The case of a frozen disordered lattice is treated by
averaging over the same set of disorder realizations
(see Ref. \cite{Fratini09}).
 In the latter case,
the results are cross-checked via an exact diagonalization of the
electronic problem on $256$ sites, 
allowing for a direct verification of
Eqs. (\ref{eq:relation}) and (\ref{eq:relinv}).

Fig. \ref{fig:theory} shows the  electron spread $\Delta x(t)\equiv
\sqrt{\Delta  x^2(t)}$  and the
time-dependent diffusivity $D(t)=\frac{d\Delta x^2(t)}{2dt}$ 
for a representative choice of
microscopic parameters:  $\lambda=0.25$, $J=0.11 eV$,
$T=0.235J=300K$, and different values of $\omega_0$.  
The dashed line in Fig. \ref{fig:theory}a is the result for static
disorder, $\omega_0=0$, showing  a finite localization length 
$L=3.6a$.
The corresponding $D(t)$ in Fig. \ref{fig:theory}b 
increases at short times in the ballistic regime,
then exhibits oscillations. 
At
subsequent times the oscillations are damped and $D(t)$ steadily
decreases and vanishes. 
Following Eq.  (2), $2(dD/dt)=C(t)$ is
precisely the velocity correlation function: 
a negative slope is therefore signalling the occurrence
backscattering  underlying the phenomenon of Anderson localization
(in a classical picture, $C(t)<0$  implies that the velocity at time $t$ is
opposite to its value at time $t=0$). 
This occurs, as expected,  at times greater than 
the elastic scattering time \cite{LeeRMP}, which is given by
$\tau_{el}=(\pi \lambda T)^{-1}=5.4$ in the present units
\cite{Fratini09}. 

The results in the presence of lattice dynamics
($\omega_0\neq 0$) closely follow the localized behavior
at short and intermediate times. 
However, upon reaching the timescale of lattice
vibrations, $1/\omega_0$ (indicated by arrows), 
localization is destroyed and 
$\Delta x(t)$ starts increasing indefinitely.
The existence of a transient localization phenomenon at
times $\tau_{el}\lesssim t\lesssim 1/\omega_0$ is one of the main
results of this work. It indicates that 
the electronic transport mechanism is 
markedly non-semiclassical, the final outcome being 
determined by the characteristic timescale  of  lattice disorder.


As a side remark, 
Fig. \ref{fig:theory}b illustrates a fundamental drawback 
of  the Ehrenfest method, that makes it inadequate to determine 
asymptotically the electron diffusion:
the diffusivity does not apparently
tend to a constant value
but rather exhibits an upward drift at long times.
The total energy of the system (not shown) is conserved in the simulation  
with a relative precision of $2 \cdot 10^{-7}$, 
which rules out possible integration errors.
The origin of this spurious phenomenon rather lies in the fact that
the Ehrenfest equations do not  properly conserve the
Maxwell-Boltzmann  statistical distribution \cite{Tully}:
the repeated action of the lattice vibrations (an
external, time-dependent potential) eventually drives the
electrons to an arbitrarily high effective temperature,  a fact that
could be at the origin of the $T^{-2}$
temperature dependence of the mobility obtained by this method
\cite{Troisi,TroisiAdvMat,Tsquare}. 

The inset of Fig. \ref{fig:theory}b shows the optical conductivity,
$\sigma(\omega)$, 
calculated by  applying Eq. (\ref{eq:relation}) to the data of
Fig.\ref{fig:theory}a, 
neglecting the spurious  superdiffusive behavior at long times. 
The result of 
the static disorder
problem ($\omega_0=0$) is shown for reference (grey, dashed).
We see that the dynamical nature of the lattice only modifies
the low frequency region
of the spectrum, $\omega\lesssim \omega_0$.
It does not affect substantially the
localization peak at $\omega_{loc}\simeq 0.4J$
as long as $\omega_0\ll \omega_{loc}$,  nor the absorption band
at higher frequencies.

\paragraph{Relaxation time approximation.}

To understand how localization features can actually coexist with 
a diffusive behavior at long times,
we now implement the relaxation time approximation (RTA) as a simple
scheme bridging between localization
and diffusion. 
The idea underlying the RTA is to express the dynamical
properties of the actual system in terms of those of a suitably
defined reference system, from which it decays over time.
Specifically, defining $C_0$ as a reference velocity correlation function,
the relation
\begin{equation}
   \label{eq:reltime}
   C_{RTA}(t)= C_{0}(t) e^{-t/\tau}
\end{equation}
describes the damping of velocity correlations caused by relaxation
processes with a characteristic time $\tau$
\cite{Mayou00,Trambly06,notedelta}.
In the semiclassical theory of electron transport,
one starts from a perfectly periodic crystal and
describes via Eq. (\ref{eq:reltime}) the
momentum relaxation due to the scattering of Bloch
states. In that case $C_0=2 v^2_{avg}$ 
is a constant (twice the thermal average of the squared band
velocity) and the resulting diffusivity, $D_{RTA}(t)=v_{avg}^2\tau
[1-e^{-t/\tau}]$, is a monotonically increasing function of time.    

One can alternatively 
take a localized system with static
disorder as the reference state, as suggested by Fig. \ref{fig:theory}.
At times shorter
than the typical timescale of the lattice motion, 
$\tau_{in}\sim 1/\omega_0$,
the molecular lattice appears to the moving electrons as an
essentially frozen, disordered landscape. In this case,
nothing prevents the buildup of quantum interferences
that lie at the origin of Anderson localization.
The velocity correlation
function $C(t)$ then initially coincides  with that of a system with
static disorder, $C_{0}(t)$. However, quantum interferences are
  destroyed at longer times because, due to the lattice dynamics,
the electrons encounter different disorder
landscapes when moving in the forward and backward directions  \cite{LeeRMP}:
Eq. (\ref{eq:reltime}) is the simplest form that is able to capture 
such decay process.
From Eqs. (\ref{eq:vv}) and (\ref{eq:relation}) it is
easy to see that, starting from a localized system (i.e. one with a
vanishing diffusion constant,
$\int_0^\infty
C_0(t)dt=0$), Eq. (\ref{eq:reltime}) restores a finite
diffusion constant,
$D_{RTA}=L_0^2(\tau_{in})/(2\tau_{in})$ which is analogous to the
Thouless diffusivity of Anderson insulators \cite{Thouless77}.   
This value is essentially equal to the diffusivity of the reference localized 
system at a time $t\approx\tau_{in}$.
The quantity $L_0^2(\tau_{in}) =\int e^{-t/\tau_{in}}
\Delta x^2_0(t)dt/\tau_{in}$ represents the
typical electron spread achieved at a time $\tau_{in}$, before
diffusion sets back in.  It therefore
acquires the meaning of a {\it transient localization length}.
The emerging physical picture is that of
electrons prone to localization,  but that can take advantage of the
lattice motion to diffuse freely over a distance 
$L_0(\tau_{in})$, with a trial rate $1/\tau_{in}$. 

From Eq. (\ref{eq:relation}) and Eq. (\ref{eq:reltime})  we obtain a
mobility
\begin{equation}
   \label{eq:dynloc}
   \mu(T)\simeq \frac{e}{k_BT}\frac{L_0^2(\tau_{in})}{2\tau_{in}}.
\end{equation}
Although a systematic  study of the electron mobility of organic
semiconductors is beyond the scope of this work, 
we note here that under quite general assumptions, Eq. (\ref{eq:dynloc})
implies  a power-law temperature dependence,  
$\mu\sim T^{-\alpha}$,  even though the microscopic transport mechanism
is far from conventional 
band transport (the exponent $\alpha$ depends on how the
transient localization length varies with the thermal lattice disorder).

  \begin{figure}
   \centering
   \includegraphics[width=8cm]{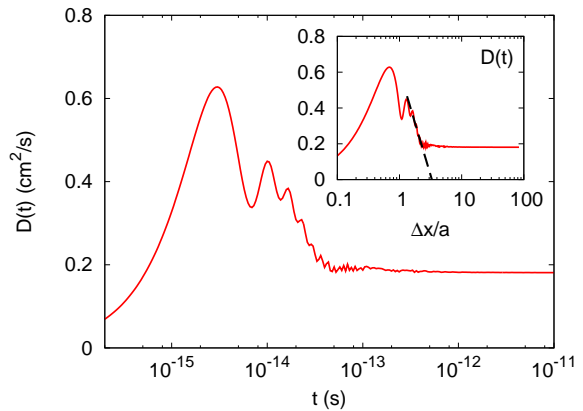}
   \caption{Color online. Time dependent
     electron diffusivity $D(t)$ extracted via
     Eq. (\ref{eq:relation}) from the
     experimental optical conductivity of Ref. \onlinecite{Basov} in
     the direction of highest conduction. The absolute value is fixed by the 
     measured mobility $\mu\simeq 7 cm^2/Vs$. 
     The inset shows the same quantity as a
     function of the instantaneous electron spread. 
     The dashed line is the weak localization
     extrapolation.}
   \label{fig:exp}
\end{figure}

\paragraph{Real time dynamics from experiment.}

We now show how optical conductivity experiments can provide direct
information on the relevant time and length scales of the problem.
Fig. \ref{fig:exp} reports the instantaneous diffusivity, $D(t)$, 
in the direction of highest conduction
of rubrene, obtained via Eq. (\ref{eq:relinv})
by direct integration of the data of Ref.
\onlinecite{Basov} (analogous results are obtained from
Ref. \onlinecite{Fischer}).
$D(t)$ increases first, reaches a maximum and
then decreases by a factor of $3$ before stabilizing to a constant value. 
The shape of the 
diffusivity curve is remarkably similar to the theoretical result of
Fig.\ref{fig:theory} and hardly compatible with the semiclassical picture
discussed after Eq. (\ref{eq:reltime}) in which $D(t)$ increases and
directly saturates. 
This suggests that, as in the model calculation, 
the decrease of $D(t)$ is due to localization effects occurring at times
shorter than the lattice dynamics. 
An elastic scattering
time of the order of $\tau_{el}\approx  10^{-14} s$ can 
be tentatively 
identified with the region of negative slope in Fig. \ref{fig:exp}.  
According to the arguments given above,
a diffusive regime [$D(t)=$constant] sets up at timescales 
beyond the inelastic scattering time. This is what Fig.
\ref{fig:exp} reveals, with
$\tau_{in}\approx 5 \cdot 10^{-14} s$.

The inset shows a plot of $D(t)$ as a function of $\Delta x (t)$
and gives access
to the relevant length scales. We find for the elastic and inelastic
mean-free paths $\ell_{el} \approx a$ and
$\ell_{in}\approx 3a$ respectively,
with $a=7.2\AA$ the intermolecular distance. In addition
this inset suggests that without inelastic scattering, i.e. for a
fixed set of disordered
molecular positions, the diffusivity $D(t)$ would extrapolate
to zero at a localization length of the  order of $3-5$ intermolecular
distances. Note that the linear extrapolation of diffusivity with the
Log of the length is a standard approximation for two dimensional
systems \cite{LeeRMP}. Yet rubrene should be considered as 
intermediate between one and two due to its highly anisotropic character
\cite{Machida}. 
For a one-dimensional system localization sets in more
efficiently, therefore the above extrapolation should be an upper bound
to the true localization length.

\begin{figure}
   \centering
   \includegraphics[width=8cm]{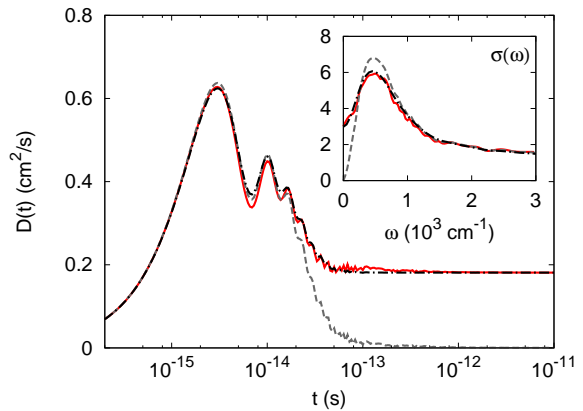}
   \caption{
     Color online. The experimental diffusivity of
     Fig. \ref{fig:exp} (red, continuous) is
     compared with the result of the relaxation time approximation (RTA)
     (black, dash-dotted) and the localization ansatz defined in the text
     (grey, dashed). The inset shows the experimental optical
     conductivity of Ref. \onlinecite{Basov} together with the RTA
     and the localization ansatz (arb. units).
}
   \label{fig:RTA}
\end{figure}

We finally show in Fig. \ref{fig:RTA} 
how the RTA can be used to extract quantitative microscopic
information from the optical data.
One first  constructs an ansatz  for the reference conductivity  
$\sigma_0(\omega)$  
representing the ideal case with
frozen disorder, i.e. no inelastic scattering.  This can be done 
starting from the experimental optical absorption (inset: red, full line),
by enforcing 
the condition $\sigma_0(\omega\to 0)=0$ appropriate for a
localized system  (grey, dashed line).
The RTA result  (black, dash-dotted) is then obtained by
applying Eq. (\ref{eq:reltime}) to fit the experimental curve. 
The optical conductivity and the resulting diffusivity
both nicely agree with the experimental data.
The fitting procedure yields $\tau_{in}
=5.1 \cdot 10^{-14}s$, corresponding to a
frequency $\omega_{in}=104 cm^{-1}$, consistent with the
relevant intermolecular phonon frequencies in rubrene
\cite{Ren09,Girlando10}. From the same fit, the estimated transient
localization length in the direction of highest conduction is
$L_0(\tau_{in})\simeq 2 a$ [the  localization length of the
static ansatz is  $L_0(t\to \infty)\simeq 3 a$].
The present analysis shows that  the finite frequency absorption peak 
observed in rubrene \cite{Fischer,Basov}
should be ascribed to transient localization effects ---i.e.  occurring 
before the dynamics of the lattice set in --- and constitute a signature 
of an unconventional transport mechanism.

\paragraph{Concluding remarks.}

The relation between the quantum dynamics of electrons and the 
optical conductivity that stems from the Kubo formula, 
appears to be a powerful tool to
analyze the charge dynamics in semiconductors 
with unconventional transport properties.
When applied to experimental data on crystalline
organic semiconductors,  it 
provides evidence for 
the  role played by localization
phenomena in the charge transport mechanism. 
The scenario emerging from the above analysis is indicative of a
prominent role of the dynamical lattice disorder,
which is supported by a microscopic calculation on a one-dimensional model.


\paragraph{Acknowledgements.}
S.C. acknowledges useful discussions with  Sara Bonella.


\begin{thebibliography}{99}
\bibitem{notedelta} Eq. (\ref{eq:reltime}) is equivalent to performing a
  convolution of $C_0(\omega)$ with a lorentzian of width $1/\tau$,
  not a convolution of the conductivity $\sigma_0(\omega)$ as is often
  done in the literature.
\end{thebibliography}
\end{document}